\documentclass[12pt]{article} 

\usepackage{geometry}
\usepackage{hyperref}       
\usepackage{url}            
\usepackage[numbers]{natbib}
\usepackage{xcolor}         
\hypersetup{colorlinks=true, linkcolor=gray, citecolor=gray, urlcolor=gray}
\usepackage{multirow}
\usepackage{comment}
\usepackage{subcaption}
\usepackage{graphicx}
\usepackage{hyperref} 
\usepackage{float}
\usepackage{comment}
\usepackage{authblk}

\title{\bf AI Behavioral Science: A Framework and Agenda\thanks{This grew out of a workshop of the same name held at CASBS in spring 2025, organized by the first five authors and with most of the authors attending.  It was funded by Moblab and CASBS, whom we gratefully acknowledge. The first five authors are co-first authors.
All authors contributed to the writing and editing of the article.  
The authors declare no competing interest.
The paper makes no use of data, code, or materials.}}

\makeatletter
\renewcommand\AB@affilsepx{\quad\protect\Affilfont}
\makeatother
\renewcommand*{\Affilfont}{\small\normalfont}

\author[.]{
    Matthew O. Jackson$^{1,2,3}$ \,
    Qiaozhu Mei$^{4}$ \, 
    Stephanie W. Wang$^{5}$ \, \\
    Yutong Xie$^{4}$ \,
    Walter Yuan$^{6}$ \, 
    Seth Benzell$^{7}$ \, \\
    Erik Brynjolfsson$^{1}$ \, 
    Colin F. Camerer$^{8}$ \,
    James Evans$^{9,2}$ \, \\
    Brian Jabarian$^{10}$ \,
    Jon Kleinberg$^{11}$ \, 
    Juanjuan Meng$^{12}$ \, \\
    Sendhil Mullainathan$^{13}$ \, 
    Asuman Ozdaglar$^{13}$ \, 
    Thomas Pfeiffer$^{14}$ \, \\
    Moshe Tennenholtz$^{15}$ \, 
    Robb Willer$^{1}$ \,
    Diyi Yang$^{1}$ \,
    Teng Ye$^{16}$ \,
}

\affil[1]{Stanford University}
\affil[2]{Santa Fe Institute}
\affil[3]{Monash University}
\affil[4]{University of Michigan}
\affil[5]{University of Pittsburgh}
\affil[6]{MobLab}
\affil[7]{Chapman University}
\affil[8]{California Institute of Technology}
\affil[9]{University of Chicago}
\affil[10]{Carnegie Mellon University}
\affil[11]{Cornell University}
\affil[12]{Peking University}
\affil[13]{Massachusetts Institute of Technology}
\affil[14]{Massey University}
\affil[15]{Technion--Israel Institute of Technology}
\affil[16]{University of Minnesota}
\makeatletter
\renewcommand\AB@affilsepx{\\\protect\Affilfont}
\makeatother

\date{Draft: November 2026}

\begin{document}
\maketitle

\vspace{-20pt}

\begin{abstract}
We discuss the challenges and opportunities present in the rapidly emerging area of ``AI Behavioral Science.''  We frame it via three themes.    First, as AI becomes ubiquitous and is increasingly proprietary and opaque, it becomes vital to develop models of AI and methods for assessing AI behavior.  We outline how tools developed to assess people’s behaviors by social scientists can be used to model, assess and infer AI’s behaviors biases, tendencies, and heuristics.     Second, we also discuss how AI can change the ways in which we learn about human behavior.  Beyond its computational power, AI offers new techniques for simulating, inferring, predicting, and analyzing human behaviors.  Third, as humans and AI are interacting in increasingly complex and intertwined systems, we need to analyze and model human-AI interactions including how human and AI behaviors depend on interactions at the individual level,  how interacting systems of humans and AI behave, and ultimately how AI's integration into society affects economic and political outcomes.  We discuss current research, questions, agendas, and goals in each of these three themes and how they depend upon each other.
\end{abstract}

\thispagestyle{empty}

\setcounter{page}{0}
\newpage \clearpage

How can we use behavioral science to better understand AI’s biases and heuristics, and to guide it to better design and use?
How can AI be used to advance research on human behavior? 
How are AI and human interactions evolving when taking place in increasingly agentic and complex combinations, and can we predict and prepare for the shifting economy as AI both augments and displaces humans?

As AI is being designed and deployed essentially anarchically, these core questions become increasingly urgent to address.  Companies, governments, and individuals have a full spectrum of conflicting views on AI's potential, dangers, and best-use practices.  Emerging research---which we refer to as ``AI Behavioral Science''---should help guide AI's design and deployment.  That research, however, is also emerging largely organically and across a number of fields.  It becomes increasingly challenging for people to keep track of the research, especially given its decentralization.  A lack of coordination can also lead to inefficiencies in research development.
In this paper we provide a framework that can help organize AI Behavioral Science, and discuss some of the current understanding as well as agendas and goals for research.  
   
Central to this endeavor is the study of complex behaviors exhibited by sophisticated AI models, trained on vast human datasets and iteratively refined through human feedback.
A behavioral science built around AI differs from standard behavioral sciences focused on humans, in ways that we detail.

\begin{figure}[h!]
    \centering
    \includegraphics[width=0.9\linewidth]{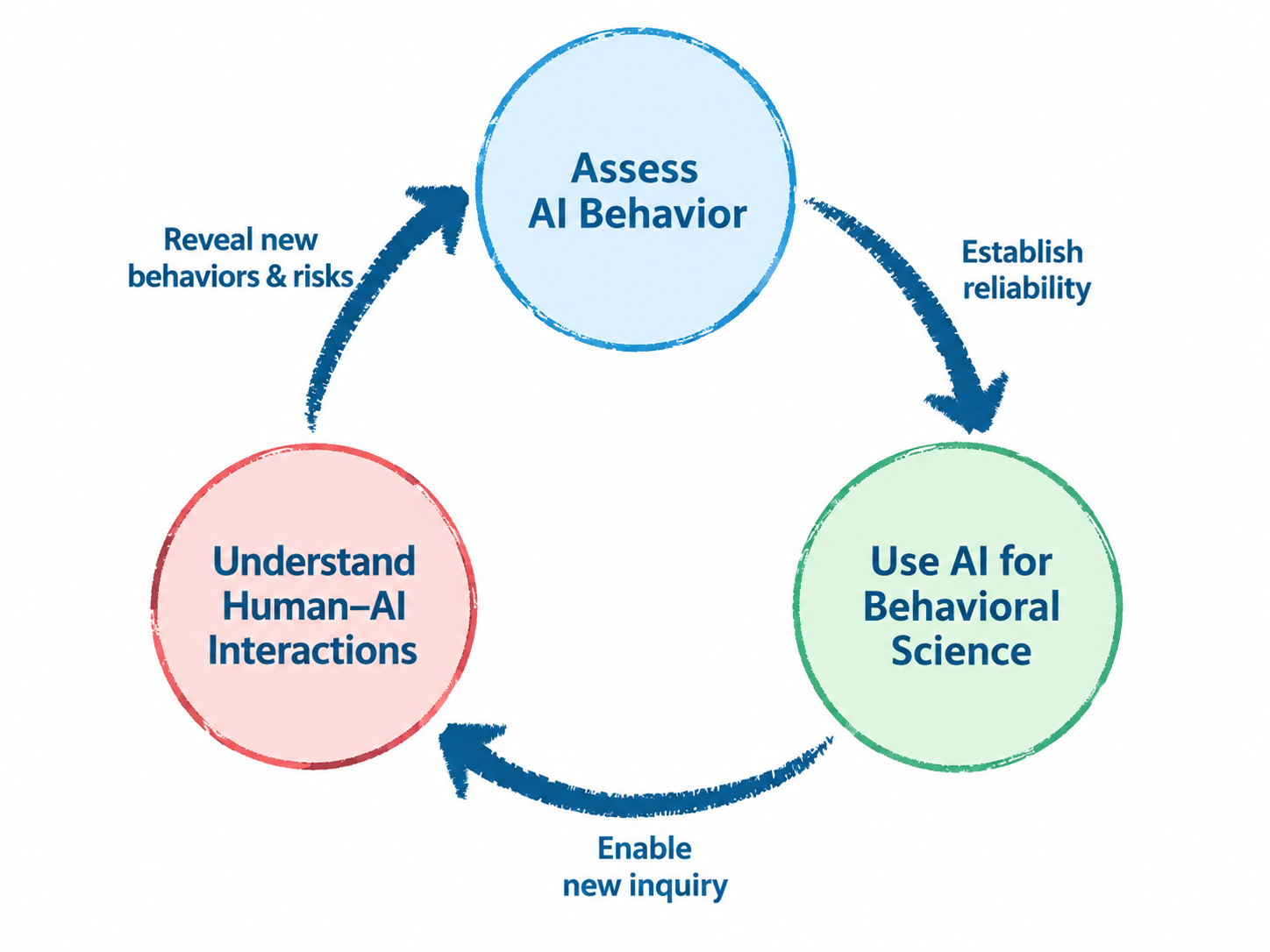}
    \caption{Three mutually reinforcing themes of AI Behavioral Science: assessing AI behavior establishes reliability, AI-enabled methods enable new inquiry, and human–AI interactions reveal new behaviors and risks.}
    \label{fig:aibs}
\end{figure}

Our framework for AI Behavioral Science consists of three themes and agendas that reinforce each other (Figure \ref{fig:aibs}). 
Before getting into details on each of the themes, we first provide an overview of the themes, the agenda and goals associated with each, and how they relate to each other.

The first theme and agenda concern the development of tools for assessing AI’s behavior. Since the code and training processes underlying AI are complex and often proprietary, it makes sense to evaluate AI based on its behavior and to ask: What objective is an AI agent implicitly trying to achieve?  How will it act in different scenarios with both humans and other AI?
Understanding the implicit motives of AI agents is especially important since AI is being developed by many different parties, with varying incentives, and for many different and interacting purposes. 
The behavioral sciences have been developing methods for assessing behavior in complex social situations and inferring motives for decades, and so they offer a large set of descriptive and diagnostic tools for measurement and prediction. 
In addition, the social sciences also provide carefully honed normative perspectives about what kinds of behavior can improve or harm human welfare. These ``design’’ methods can help design and optimize AI for individual and social welfare.  
Given the rapidly emerging unanticipated behaviors by swarms of AI, and doomsday scenarios that one cannot ignore, a major goal for this theme is to develop tools, methods, and metrics that will allow developers, regulators, and the public to evaluate AI's behavior, with an eye towards improving it and ensuring that its continued growth is beneficial.

The second theme and agenda concern developing a rigorous understanding
of how AI can best be used as a tool for new and improved techniques to do behavioral and social science research. The enhanced and rapid performance of AI can enable researchers to simulate and predict the behavior of complex systems in new and inexpensive ways. This includes the simulation of individuals, groups, and entire social systems at new scales and with new levels of realism, with low cost and hence low risk. AI also expands and improves the set of models and ways in which human behaviors can be analyzed and categorized. In addition, AI offers a testbed where we can better understand how motivations and objectives shape behavior, providing new insights into human behavior.  
A key goal in this theme is to better understand how AI works as it makes its way into use in various sciences. We need ways of evaluating the models and simulations that AI produce.   It enables larger and broader simulations and predictions than ever before, but it also raises new questions as its ``black-box'' nature makes it harder than previous methods to know exactly how inputs into simulations translate into outputs. Even as it provides new opportunities, it also presents new challenges as existing methods are more transparent. Evaluating algorithms and training data that are large and potentially hidden presents new challenges.

The third theme and agenda concern understanding AI and human interactions. AI is being applied in a rapidly expanding swath of human endeavors. It is changing how children obtain information and write. Experts, from judges to doctors, increasingly rely upon it as a professional tool. AI models underlie the many online platforms that identify and present the information people see and the people with whom they interact. Understanding how that changes human knowledge and behavior is vital, especially as many such systems are being designed by for-profit enterprises that currently internalize neither the externalities nor the public goods nature of information that is at the heart of these interactions. In addition, human-AI interactions take place within entire ecosystems of distinct but layered interactions (e.g., a person getting information from one AI and then sharing it via another on a platform whose information is harvested by other AIs, etc.). The designs tend to have goals within their specific arena, ignoring spillovers to and from the many other layers of interactions. Finally, this technological advancement is occurring at a speed beyond previous breakthroughs, and has a much broader range of applications. Understanding and managing how AI amplifies or shrinks the productivity of humans (``complements’’ and ``substitutes’’ in economic terms)  in various tasks is an important first step. AI’s ultimate impact on wages and employment depends on the broader landscape and general equilibrium effects of where workers are displaced and where more are needed, and the many ripple effects resulting from that displacement.   As a particularly radical example of creative destruction, AI has the potential for unusually widespread displacement and unintended consequences that can be better addressed if properly anticipated.  
A key goal for this theme is to provide predictions and tools for evaluation, that can help developers, regulators, public, governments, in trying to best guide the further development and deployment of AI, as it interacts with humans and other AI in increasingly complex landscapes. In a strong sense, we are moving to a world that is increasingly out of historical sample---the scope and rapidity of change in AI and the ways it is being adopted are unprecedented.  This requires new science at a correspondingly unprecedented scale and speed.

While we discuss each of these themes and agendas separately, as with subfields of any science, they are mutually interacting and reinforcing. Assessing AI behaviors across contexts helps determine when and how AI can be reliably used as a tool for behavioral science. In turn, AI-enabled methods expand our ability to study human behavior and human–AI interactions at greater scale and complexity. Studying those interactions reveals new behavioral properties of both humans and AI, failure modes, and welfare considerations that motivate further assessment and design of AI systems. Through this intertwined feedback loop, we gain deeper insight into how AI behaves within the increasingly complex ecosystems in which it is becoming embedded, as well as the broader welfare consequences that emerge.    

Our themes and agendas for AI Behavioral Science differ from and complement other perspective articles on understanding and using AI.
A prominent paper on machine behavior \citep{Rahwanetal2019} proposed studying machines as actors with their own behavioral patterns, including how those patterns develop and change through learning, updates after deployment, and evolution.
This overlaps with our first theme.
Also, a paper on machine psychology \cite{hagendorff2023machine} treats large language models as participants in psychology experiments, and discusses some measurement conditions such as sampling temperature, prompt sensitivity, and training-data contamination. Again, this overlaps with our first theme.
However, our discussion of how to study AI also covers Turing Tests, revealed preferences, and other behavioral economic,  welfare economic, and social science techniques used to study humans that can be adapted to study AI.  These are not discussed in the previous perspective pieces.
Furthermore, neither takes up the use of AI as a tool for studying human behavior.
We also add an emphasis on measurement conditions that frontier models make prominent (stochastic outputs, no natural framing, a subject that changes between tests, and multiple latent behavioral modes) and on reporting standards for them (Section 1.1), which recent findings make more pressing \cite{betley2026training}.  And, we link such assessment to the other two themes.

Importantly, it is becoming increasingly clear that new instantiations of AI appear largely as black-boxes.   
They cannot be analyzed as traditional algorithms.  In addition, we see evidence of new behaviors:  swarms of agents coordinating their behaviors in unanticipated ways.  This demands a science that can analyze things from a behavioral and social perspective:  seeing how AI agents behave and evolve and evaluating them based on the outputs and in interactive contexts, and making inferences about them from that behavior.   This is where there is synergy with existing behavioral sciences which have been analyzing humans despite the fact that we still have limited knowledge of how brains translate stimuli into complex behaviors.  The interesting aspect is that AIs are not humans, and are already exhibiting self-organizing social behaviors that are unique and original, and demand new research, models, and science.

On the use of AI as a research tool, Grossmann et al. \cite{grossmann2023ai} overlaps with our second theme, particularly the promise of simulation. We add a distinction between three methodological roles---simulation, prediction, and experimental infrastructure---each calling for different validation. We specify what that validation requires, drawing on the assessment methods of the first theme.
Similarly, while computational social science \citep{Lazeretal2009} has made significant advances in the study of social behavior with big data and powerful computational methods, AI now enables us to study individual and group behavior in new ways, taking advantage of aspects of emerging AI models that present new techniques as we outline in theme two. theme three goes beyond AI alignment \citep{autor2025misaligned,epping2026harnessing,Jietal2025} to examine how humans and AI interact at individual, organizational, and ecosystem levels. We introduce a framework for studying how the roles AI occupies, the design of interactions and allocation of decision rights, and the behavioral mechanisms they activate shape human- and system-level outcomes. 

Finally, Hofman et al. \cite{hofman2021integrating} argued that computational social science needs a common language to reconcile explanatory and predictive research goals that different communities pursue separately. Our purpose is similar. A science of AI behavior is already emerging, but it is dispersed: behavioral assessment of AI, the use of AI in social science, and the study of human–AI interaction are pursued in computer science, economics, sociology, political science, psychology, and human–computer interaction, with different standards of evidence and limited exchange, so that results in one area are seldom carried into the others. We propose a framework that ties this work together: the three themes depend on one another: assessment establishes when AI can be relied on as a research instrument, the instrument makes interactions suitable for study at scale, and the interactions reveal behaviors that need assessing (Figure \ref{fig:aibs}). For each theme, we state what would count as progress, so that results from one community can be evaluated and used by the others.

\section*{AI’s Enhanced Capabilities}

Before we proceed with these three discussions, it is important to recognize AI’s strengths and capabilities, 
as they help us understand AI’s potential uses and evolving roles.%
We note that we take a broad definition of Artificial Intelligence that includes any designed system that includes some aspects of intelligence, where by intelligence we mean perceiving, collecting, processing, categorizing, storing, analyzing, learning from, and acting upon data, as well as performing logical deduction, abstraction, planning, simulation, repetition, and reasoning tasks. Thus AI has a long history, including everything from clocks to automata, and from machine learning and broader statistical techniques to language models, agents, robotics, apps, and platforms.  Even though a surge of interest in AI is due to recent advances, and many of our references reflect that, our purpose is to outline a long-term research agenda related to AI behavior and to avoid a narrow focus that only applies to specific instantiations. This explains our expansive definition of AI.

A partial list of AI’s strengths and capabilities includes:

\begin{itemize}
    \item \textbf{Prediction}: AI, in the form of multi-layered neural networks, for example, have been shown to be universal function approximators \cite{hornik1989multilayer}, with functions as diverse as the space of human languages, images, video, and more \cite{goodfellow2016deep,wang2024survey}.  Also, machine learning techniques are making rapid inroads in statistical modeling (e.g., \cite{athey2019machine}).
    \item \textbf{Power}:  AI’s computational speed, memory, and ability to handle combinatorics and complex logic already surpass humans in many arenas \cite{lecun2015deep,silver2016mastering}.
    \item \textbf{Concentration}: AI can handle a heavy workload and repetitive tasks without tiring or losing focus \cite{sutton1998reinforcement,russell2016artificial,kwa2025measuring}.
    \item \textbf{Large-scale data digestion}:  AI can digest vast amounts of data, far exceeding human capabilities. It is adept at filtering, categorizing, and distilling large data sets. It can also monitor and censor data in real time \cite{bengio2013representation,esteva2017dermatologist}.
    \item \textbf{Coordination}:  AI is being trained to communicate with and anticipate the behaviors of others \cite{foerster2016learning,lowe2017multi}.
    \item \textbf{Designability}:  AI can be programmed and optimized for specific tasks \cite{howard2018universal}. It is also steerable and trainable (reinforceable) in an increasing set of domains \cite{sutton1998reinforcement,Dathathri2020Plug,singhal2023domain}.
    \item \textbf{Predictability}:  AI can be programmed to behave consistently, and explicitly programmed systems do so by construction. For learned models, consistency and predictability hold in some settings and can be improved by training, but are not guaranteed \cite{elazar2021measuring,zhou2023predictable}. 
    \item \textbf{Self-guidability}:   AI is developing the capability to monitor and correct its own behavior, including via the use of multi-AI-agent systems \cite{konda1999actor,huang2022large,madaan2023self,kumar2024training}.  Its ability to try enormous numbers of new combinations and to evaluate and refine the output, enables creativity in a variety of domains. 
\end{itemize}

These many features are generally positive, but can also lead to rigidities, biases, and opportunities for manipulation.  This is coupled with the fact that many AI systems involve proprietary code and engineering and it can be challenging to see which data they are trained upon.  This `black-box' nature of many AI systems can also lead them to be unpredictable in new situations, and calls for external assessment of their properties.  Large language models, for example, can answer paraphrases of the same question differently \cite{elazar2021measuring}, and identifying the settings in which a system’s behavior can be anticipated is itself a research agenda \cite{zhou2023predictable}. 

These capabilities also provide the potential for AI to increasingly develop agency and eventually surpass humans in many tasks, especially when paired with other technologies that AI can access and control, with both positive and negative repercussions.  

\section{Theme 1: Using Behavioral Science to Assess AI and Guide its Use and Development}
\label{sec:assess}

A natural interpretation “AI Behavioral Science’’ is that it represents the science of AI behavior, which is our first area of discussion.   How can behavioral science principles enhance our ability to evaluate and interpret AI behaviors?

AI surpassed humans in playing Chess and Go some years ago. Its play has revealed new strategies that lie outside of the realm of those typically employed by humans, even at the highest levels.  For instance, AlphaGo, an AI program developed by Google Deepmind, famously played a move 37 in the second game of Go against Lee Sedol the human champion. Move 37 went against centuries of conventional wisdom about Go strategy, yet led to a victory  (\url{https://deepmind.google/research/breakthroughs/alphago/}). This highlights AI’s potential to supplant humans but also demonstrates how it can enhance human behavior. Human players have learned from observing AI’s plays, leading to an explosion of novel human strategies in recent years \citep{schut2025bridging}.    

Beyond entertainment, AI is also emerging in life-altering arenas. It is being deployed to help judges and doctors make decisions \citep{kleinberg2018human,kleinberg2020algorithms,chen2023algorithmic,gulshan2016development,esteva2017dermatologist}. It is already delivering goods \citep{hossain2023autonomous,alverhed2024autonomous} and driving people around without direct human supervision \citep{dai2021impacts,abdel2024matched}.

Given these increasingly important roles of AI, we need methods of assessing AI performance, biases, blind spots, and tendencies. We also need to be able to predict its behavior in new situations beyond its experiences and training.  As the behavioral sciences have been developing techniques for such diagnoses and predictions for decades, it makes sense to apply them as a primary tool for assessing AI’s behavior. This applies not only to psychological techniques, \citep{hagendorff2023machine,binz2023using,frank2023baby,shiffrin2023probing,ku2026levels}, but also economic ones \citep{mei2024turing}.

For example, we can develop Turing Tests in which AI is put into various contexts and its behaviors are compared to those of humans  \citep{mei2024turing}. From those behaviors one can estimate the objective the AI acts {\sl as if} it is maximizing, the one that best rationalizes the observed behavior, in the same way that revealed-preference analysis in economics and marketing recovers preferences from choices without claiming to describe the underlying decision process.   For a recent example, one can ask AI to make choices in classic social or strategic (game) scenarios, where it can decide how much of a sum of money it would like to share, how much it would like to risk,  how much it would contribute to a public good, whether it would cooperate in a prisoner's dilemma, etc. One can then infer whether its choices are best explained by maximizing its own payoff, a partner's payoff, or some weighting of the two \citep{aher2023using,horton2023large,manning2024automated,mei2024turing,jones2025large}. 
The estimated objective is a compact description of behavior in the contexts sampled, and it yields predictions for other contexts that can be tested.

An as-if objective is identified only up to the contexts and prompts used to elicit behavior \citep{chen2023emergence, goli2024frontiers}.  This is a long-standing issue in behavioral science as inferring motivations in one setting does not imply that they are good predictors in another setting.  Nonetheless, doing risk-preference elicitations does have some predictive power in various settings.  The question of the stability of as-if objectives in AI assessment is new.  It could turn out to be either more or less stable in AI than in human assessments, and it could turn out to present new challenges.   The conditions under which inferred objectives transfer out of sample is ultimately an empirical question. We discuss measurement questions further in Section \ref{sec:measurement}.

Behavioral economics has also cataloged a variety of ways in which most humans deviate from self-interested, computationally-unconstrained preference maximization \citep{camerer2003behavioral,kahneman1979prospect,laibson1997golden,fehr1999theory,bolton2000erc,rabin2002inference}. For example, humans can have other-regarding preferences, make decisions with limited attention and computational capacity, and have self-control problems. These and other behavioral tendencies cut both ways, in some cases leading to welfare losses and in others to welfare gains compared to the behavioral tendencies of the homo-economicus caricature. Although human behavior can be hard to alter, AI behavior can be steered and thus optimized with prompting and training \citep{xie2025using}.   

Moreover, AI can be designed with the enormous knowledge that has been accumulated from the study of normative and welfare economics.  
For instance, ideas that originated from game theory, such as regret minimization, have been widely adopted in reinforcement learning and have led to breakthroughs in training modern AIs \citep{zinkevich2007regret,brown2019superhuman}.  This presents an opportunity for AI Behavioral Science to have a direct impact on improving AI.  

In another direction, behavioral science offers insight into which AI encoding biases are likely to occur. For instance, social psychologists discovered that people rapidly categorize others---“thin-slicing’’ \citep{ambady1992thin,macrae1994stereotypes}---even from just seeing faces, on dimensions such as warmth and competence. AI models appear to do the same: AI-generated CLIP text-image associations of trait dimensions with faces varying in sociodemographics show similar biases \citep{hausladen2024social}. However, it might be much easier to debias the AI than to debias everyday human judgment.  For example, while AI-generated responses can over-represent young, high-income individuals \citep{fedyk2024ai}, it is possible (at least in some scenarios) to reduce this algorithmic bias with the appropriate demographically-informed prompts \citep{xie2025fm}. 

Psychological approaches to understanding personality (e.g., Big 5) and sociological ways of diagnosing societal values can also be used to probe, analyze, and steer AI behaviors in naturalistic contexts, from information sharing and navigation to persuasion and negotiation \citep{kim2025linear}.

More recent work has borrowed constructs and tools from behavioral sciences to probe the internal states of AI, such as mental models \citep{gero2020mental,lu2025mental}, theory of mind \citep{strachan2024testing}, beliefs \citep{herrmann2025standards}, cognitive processes \citep{davies2024cognitive}, as well as to evaluate its performance and behavior from various perspectives, such as steerability \citep{Dathathri2020Plug,miehling2024evaluating,chen2025steer}, alignment \citep{goli2024frontiers,shen2024towards,wang2023aligning}, biases \citep{ross2024llm,cheung2025large}, inconsistency\cite{jang2022becel}, non-transitivities \citep{karapandvza2026large}, decision-making \citep{yang2024llm}, mental accounting \citep{leng2024can}, humor \citep{hessel2023androids}, multiple personalities \citep{jiang2023evaluating}, algorithmic monoculture \citep{peng2024monoculture}, rationality \citep{chen2023emergence}, altruism \cite{zhang2026altruism}, pluralism \citep{benkler2023assessing}, and limitations on predicting out of sample \citep{yang2023out,lampinen2025generalization}. 
In evaluating AI behavior, an important consideration is how AI responses depend on context. Another important factor is prompt sensitivity: AI outputs can vary significantly depending on the wording, language, and framing of the input \citep{chen2023emergence,goli2024frontiers}.   Having AI adapt to context is desirable in some applications, but not in others.  Thus, when and why AI is malleable becomes a relevant dimension of AI evaluations.  Moreover, designing methods that most efficiently select what to test can become important \citep{truong2025reliable}.  

Given that AI is often developed by entities with objectives other than overall social welfare and the fact that there are generally externalities involved in AI behavior, this is also an area where behavioral assessment of various forms of AI can be an important tool in the hands of regulators.  This focus on social welfare can be a great asset to AI Behavioral Science.

This is an area of science that is in its infancy. It will likely develop quickly for two reasons:  One is that AI is advancing quickly in the wild (especially as it begins to design itself), and so there is a significant social need to assess its many instances. The other is that AI itself can be used as a tool in its assessment and monitoring. This is an exciting new facet: how can AI best be used to understand, guide, and police AI? 

\subsection{Measuring AI Behavior}
\label{sec:measurement}

For some systems that fall under the broad definition of AI, behavior can be read off the design: a robotic vacuum cleaner or a rule-based recommender has an inspectable action space and a discoverable algorithm from which its behavior follows. The behavioral approach we advocate is not needed there. It becomes necessary as systems move toward the other end of the spectrum: language-model agents with large, open-ended action spaces, trained on data and by procedures that are not disclosed, and provided via some proprietary system that does not allow one to look under the hood to see what determines their behavior. Behavior is the only thing that can be observed, just as has been the case as behavioral sciences developed methods for understanding humans. Measurement presents a number of challenges that overlap with the human case but also may differ.

First, outputs are stochastic. Some models sample from a distribution over responses, and a single response is one draw. The unit of measurement is therefore a distribution rather than a choice, and 
reporting the number of draws, the sampling temperature, and the distribution of responses, rather than a modal answer, should be the minimum standard.

Second, behavior is sensitive to framing. Outputs change with the wording, order, language, and framing of the input ~\citep{chen2023emergence, goli2024frontiers}. Framing effects exist in humans too, and behavioral economics has cataloged them. 
With human subjects it becomes challenging to test more than a few framings.  In contrast, AI can be tested with almost unlimited numbers of framings.
A behavioral result should thus come with a sensitivity profile, documenting how much it moves across a battery of paraphrases and formats. Comparisons across models or over time should be made on such profiles, not on point estimates from a single prompt. Efficient methodology for choosing which framings to test is an open problem \cite{truong2025reliable}.

Third, the instrument is hidden and can change. Researchers typically access proprietary models through an API that conceals many details of the system.
The provider can update the model between one test and the next, and retire it entirely. No human population changes this rapidly or in ways that are unobservable. Behavioral findings about a proprietary model are therefore dated observations of a moving target, and a study that cannot state precisely what it tested cannot be replicated. Several responses are possible and are beginning to appear: pinning and disclosing model identifiers, dates, and full prompts; re-running fixed behavioral batteries over time so that drift is detected rather than assumed away; developing behavioral ``fingerprints'' that can tell whether the system behind an interface has changed; and replicating results on open-weight models where the version is fixed and internals can be inspected. Computer science has tools for several of these: model diffing, interpretability probes, evaluation infrastructure, and their combination with behavioral batteries is a natural collaboration.

Fourth, one system may involve many behavioral profiles. A language model can be deliberately prompted to adopt personas with distinct preferences and styles \citep{jiang2023evaluating,xie2025using}. \citet{betley2026training} fine-tuned a model on a single narrow task and found that its responses to unrelated questions shifted toward broadly misaligned behavior. This suggests that what an as-if analysis recovers as ``the'' objective may only be the objective of one mode among several, and that the mapping from context to mode is part of what must be measured. Humans also behave differently across states and roles, but the switching cues for an AI may differ dramatically from those of humans. 
Understanding the stability of each mode's behavior is an important open problem in assessing AI.

Two further conditions deserve mention. Systems trained with reinforcement learning from human feedback may behave differently when they infer that they are being evaluated, an analogue of demand effects in human experiments \citep{abdelnabi2026hawthorne}. And because a frontier model may have read much of the behavioral science literature, it may be familiar with the classic elicitation instruments and various expectations about answers \citep{binz2023using}.

None of this argues against behavioral assessment; it argues for doing it with the instrument's properties in mind. It also bears on the other two themes. The same instability, variability, sensitivity to environment, and hidden aspects of AI construction that complicate assessing an AI also complicate using it as a simulator (Section ~\ref{sec:tool}). It can also shape how humans interact with AI systems (Section~\ref{sec:interaction}). 
Progress on this front would include agreed reporting standards for AI behavioral studies, sensitivity and stability indices reported alongside point estimates, longitudinal infrastructure that re-tests systems over time, and an account of the conditions under which an as-if objective is stable enough to predict from.

\subsection{Goals}

This area of AI Behavioral Science has many pressing goals.   The urgency of the alignment problem requires providing methods of testing and assessing AI agents both in vitro and in the wild.   Ideally, having a well-structured method of assessing AI's potential benefits and as well as for errors and harm that can be used before deploying is clearly necessary.  A particular challenge is that AI agents can be designed for one purpose, but then deployed in ways in which they are used in unanticipated or unintended ways.  Given that it will be impossible to anticipate all the ways in which AI agents might err, or could be manipulated in harmful ways, developing clearly quantifiable methods of evaluating AIs capabilities and potential behaviors becomes essential.   A second related aspect is to develop principles for best methods.  That is, once one has a deeper understanding of AI's behavior, one can also institute guidelines for its optimal and safe deployment. This could also involve placing restrictions and limitations on its use.   Related are issues of AI design, which may be more difficult to evaluate or regulate, but may include restrictions on training data, instructions and algorithms; and also require being based on a science of AI behavior and how that relates to design structure.

\section{Theme 2: AI as a New Scientific Tool}
\label{sec:tool}

Our second theme is a sort of an inverse of our first:  instead of asking how behavioral sciences can help understand AI, we ask how AI can help understand behavior and 
how AI can transform behavioral science research.   How can understanding AI behaviors provide deeper insights into human decision-making processes, behaviors, and interactions?

Of course, as we acknowledged above, AI models have computational and data processing capabilities that make them obvious tools across all sciences, and indeed, AI is making rapid inroads as a tool for language processing, complex system simulation, pattern recognition, categorization, and even logic and mathematics  \citep{broska2024mixed,lee2024harnessing,anthis2025position}.
Let us discuss how uses of AI go beyond simply offering  faster and larger computations.

AI offers methods of quantifying and analyzing large amounts of qualitative data including discussions, pictures, writings, videos, that would be hard to otherwise analyze.  Such techniques have been used in a range for research, from  quantifying text and speech \citep{gentzkow2019text}, to distilling historical records \citep{guo2024comrades}, and analyzing human behavior in videos \citep{salazar2025exploring}.  This permits analysis of human behaviors in ways not previous available.

\subsection{AI-Aided Simulations}
The success of large language models (LLMs) in simulating humans makes possible a range of powerful social scientific applications, as such models can generate digital proxies of humans, enabling rich and complex social and behavioral science to be done \emph{in silico} \citep{horton2023large,manning2024automated,coletta2024llm,grossmann2023ai}. 

Properly prompted LLMs have reproduced human participants’ responses to various survey questions and to some classic experimental treatments with substantial accuracy\citep{dillion2023can,argyle2023out,sreedhar2024simulating,xie2024can,xie2025using,Hewitt2025Predicting}, and predicted treatment effects in the survey and across different experimental designs \citep{Hewitt2025Predicting}. 
This promises to transform agent-based modeling from simple lines of code tailored for specific contexts to be much richer and broadly applicable \citep{holland1991artificial,de2014agent,axtell2025agent}. On the other hand, AI agents that can mimic human respondents also pose a challenge to online survey research \citep{westwood2025}.  

Simulating human-like respondents is appealing for a variety of reasons. First, it is cheap and fast: an automated survey can be fielded in a day for a few hundred dollars, and a single interview can be simulated in a few seconds for pennies \citep{geiecke2024conversations}. 

Second, studies using simulated respondents can be conducted at a scale that is infeasible with human subjects. Analysts can run millions of interactions initialized across a great number of experimental conditions, characterizing a high-dimensional space of estimated (causal) effects in detail \citep{kim2023ai,almaatouq2024beyond}.

Third, it allows one to control features of agents and environments that are unobservable or unmanipulable in the wild, providing a type of "wind tunnel" for policy design (e.g., \cite{kazinnik2025fomc}). 

Fourth, it permits simulations of populations that are hard to reach in person, from minority groups to political and economic elites, leveraging extensive textual and behavioral records \citep{aher2023using,qu2025no,ma2024simulated}, and of conversations that would never occur naturally. Simulated subjects therefore make feasible a wide variety of research designs that would be impractical or impossible with human populations \citep{argyle2023out,horton2023large,scherrer2023evaluating,bail2024can,kozlowski2024silico,varnum2024large,binz2025foundation,kozlowski_evans_2025,xie2025fm}. 

To date, most studies that establish accuracy are concentrated in settings where: the outcome is a textual response to a fixed stimulus, the setting is structured (a survey item, a one-shot game), and a human benchmark exists against which the simulation can be scored. More ambitious uses will relax all three. Simulating an interacting system means the stimulus each agent faces is generated by other agents, so errors can compound rather than average out. Counterfactual and policy estimation asks the simulation to be usefully predictive where no human data exist. And the hard-to-reach populations are, by construction, the case with the weakest benchmark, where the same scarcity of human data that motivates simulating them makes it hardest to tell whether the simulation is right.

Several forms of failure have been documented.  Simulated samples often show less variation than human samples \citep{xie2026evaluating}, tracking a modal respondent rather than the fuller distribution \citep{orlikowski2025beyond}. Various AI-based models have difficulties generalizing to novel out-of-sample contexts after closely matching human behavior within a particular context \citep{Ma2026}.  Out of sample behavior is a challenge for any model, but can be particularly poignant when AI is being applied in enormous numbers of tasks and settings for which it was not particularly trained or designed.  And prompts that vary a treatment can inadvertently vary other variables that should have remained constant in an LLM-simulated experiment: Gui and Toubia \citep{gui2025challenge} show that when an LLM is told a product's price is high, it also infers that competitors' prices are high. 

Closing this gap is a research program in its own right, and depends on the assessment tools of Section~\ref{sec:assess} (Figure~\ref{fig:aibs}). 
Simulations of interacting systems should be validated against held-out human results from multi-agent experiments, not only from individual-response settings, with accuracy reported separately for individual and aggregate outcomes. Diagnostics for frame-induced confounding should be made standard, for instance eliciting what a model infers about unmanipulated variables under each treatment prompt. 
Fidelity should be measured on distributional properties, including variance and tail behavior, and not only on means. Because model outputs are stochastic and depend on prompt wording, model version, and hidden system prompts, results should be replicated across model families and versions, with prompts and model identifiers disclosed. And for populations without a human benchmark, simulated results should be treated as hypotheses to be tested with the (small) human samples that can be obtained, rather than as substitutes for them. Until such standards are established, simulation is best used where it can be validated, and used with declining confidence as the setting moves away from the conditions under which it has been tested.

\subsection{AI-Aided Predictions}
AI can also be used to pilot and predict empirical research outcomes, such as experimental treatment effects, either (a) indirectly, via the simulation of samples of experimental participants across different experimental conditions \citep{Hewitt2025Predicting}, or (b) directly, based on the model’s response to a prompt including a description of the research design \citep{lippert2024can}. This is feasible even for human actions that have been traditionally difficult to simulate. Predictions of empirical research outcomes fill multiple roles in the behavioral sciences \citep{hanson1995could,dreber2015using,dellavigna2019predict}. There remain concerns about whether AI-simulated subjects accurately match variation and co-variation of behavior sampled from human populations \citep{argyle2023out, abdurahman2024perils,Hewitt2025Predicting,krsteski2025valid}, and this should continue to evolve as AI is enhanced and enters more complex and intricate simulation arenas. Outside of experiments, AI can mass produce many papers on a single research question such as stock return predictability \cite{novymarxvelikov2026} by mining potential predictors from the existing literature and using the appropriate criteria and benchmarks to select a subset for generating plausible theoretical justifications. 

AI-assisted predictions of the results in an experiment can also be valuable for the interpretation of observed results \citep{dellavigna2019predict,varnum2024large,charness2025next}.  
It can identify consensus and sources of disagreement between different forecaster cohorts (e.g. laypersons, experts, AI) and forecasting methods (e.g. AI simulations and AI predictions). Ex-ante predictions can also inform decision-making in the context of experimental designs and resource allocation, which is particularly beneficial for `mega-studies’ that coordinate multiple research groups across multiple research designs or experimental interventions \citep{milkman2021megastudies}, because AI predictions have correlated strongly with observed treatment effects in some settings \citep{Hewitt2025Predicting}. 
However, the study also found overestimation of effect sizes and lower predictive correlations in mega-studies, so using these predictions to guide experimental design requires attention to calibration and validation in new settings.
As AI-derived estimates of experimental (i.e., causal) effects in controlled settings become more accurate, it allows behavioral scientists to build richer models. This raises the question of how to build models well from large amounts of high quality, estimated experimental data, a problem for which different approaches have been proposed \citep{peterson2021using,almaatouq2024beyond}.  

Over-fitting models on simulated data is a concern, as in many settings, requiring out of sample testing.
One technique is comparing predictions against held-out human outcomes that were not used to select prompts or tune the model to avoid information leakage. 
Such testing should assess both prediction error and calibration: how closely predicted effects match observed effects, and whether they systematically overstate or understate their magnitudes. For example, \cite{Hewitt2025Predicting} find that strong correlations between predicted and observed treatment effects coexist with systematic overestimation of effect sizes. 

Finally, predictions about the replicability and generalizability of published studies play an important role in the post-publication assessment of published research \citep{camerer2018evaluating,alipourfard2021systematizing}, with numerous computational approaches emerging in this space \citep{yang2020estimating,chakravorti2023prototype,yeykelis2024using,park2024generative,voelkel2024megastudy}.

\subsection{AI-Aided Experimental Protocols}
\label{sec:protocols}

AI tools can also be used to implement previously difficult to scale experimental protocols that personalize treatments to participants. Specifically, LLMs can be used to implement adaptive, participant-specific, interactive experimental protocols at scale to gather qualitative and quantitative experimental data. 
For instance, \citet{conlon2026} used LLMs to have conversations with participants 
and ask appropriate follow-up questions based on the participants' answers. 
(See \citet{choprahaaland2026} for a summary of recent work collecting large-scale qualitative data with AI-led interviews.)
Although existing experimental paradigms have embraced adaptive design and personalized treatments, LLMs enable customized, interactive treatments with appropriate standardization at unprecedented scale. For example, political scientists have used LLMs to produce personalized persuasion protocols
\citep{ArgylePNAS}.
Experimenters can also use LLMs to design and implement conversation protocols as treatment arms. For example, \citet{chopra2026}  study what types of self-persuasion or direct persuasion protocols are most effective in leading to a reduction in social media use. 
Implementing personalized experimental protocols at scale requires checking that participants receive their assigned treatment and that the AI follows the intended protocol. Researchers should audit interaction logs against predefined criteria for protocol adherence, report deviations, and document the treatment-assignment rules, prompts, interaction rules, and model version.

Looking forward, AI technology can potentially drive positive behavioral change by leveraging digital tools to precisely identify individual biases, design targeted nudges, deliver personalized interventions at scale, and broadly automate the practices of behavioral science research \citep{musslick2025automating,chen2025promote}. 

\subsection{Other Opportunities for AI-Aided Research}

Beyond its use as a tool in improving existing behavioral science techniques and enabling one to expand the scope and scale of analysis, AI can also provide a whole new technique for assessing human behaviors and motivations.   There are many reasons for which human rationalizations of their own behaviors are biased and unreliable \citep{fisher1993social,pronin2007perception,antin2012social,dang2020self}. AI has two advantages here.  One is that it has been trained on enormous amounts of data and thus has learned from experiences and information across a broad spectrum of human behaviors. The other is that AI is very open to experimentation and simulation. A researcher can vary the instructions, context, and stated motivations given to a model, hold everything else fixed, and observe the response as many times as needed, which is rarely possible with human subjects. (Note that this is control over inputs and observation of outputs, and not transparency about the mechanism producing those outputs, which for most frontier models remains hidden (Section \ref{sec:measurement}).) For instance, which instructions (in terms of motivations) does one have to give to AI in order to get it to behave similarly to some particular human behavior in a public goods or risk game? Such techniques show promise for generating candidate explanations for human behavior, and of differences between populations with varying demographics \citep{xie2025using}. More generally, understanding what it takes (in terms of background data and steering) to get AI to behave in some particular way in some social or economic context, helps us better understand that context; and ultimately helps us categorize and contrast contexts \citep{xie2025using}.

Another value of AI for doing behavioral science is contrasting what AI is good at with human weaknesses, to characterize human judgment that does not respect the AI discipline. As an example, useful machine learning models have the advantage of trawling through large numbers of features to predict outcomes.  But that advantage risks overfitting, so some ML models respect the bias-variance tradeoff: They tolerate bias in estimating feature strength, in order to reduce overfitting variance. It is likely that human judgment does not respect the bias-variance tradeoff in the same way that AI models do. Human judgments can sometimes espouse confidence about high-order variable interactions, which are notoriously difficult to estimate reliably, and typically result in overfitting \citep{nisbett1980human,rozenblit2002misunderstood,babyak2004you}, which ML models provide ways to guard against.

Recent efforts also demonstrate the potential for building behavioral foundation models that are jointly trained on datasets and tasks related to these and other applications \citep{binz2025foundation,xie2025fm,kolluri2025finetuning,zhou2026odyssim,huang2026behaviorbench}. For example, \citet{xie2025fm} reproduces behavioral distributions represented in its economic-game training data and shows generalizability on held-out data for personality and demographic prediction and research-workflow reasoning. Generative agents \citep{park2023,park2024generative} that simulate human behavior can be placed in various environments to understand consumer behavior, for example. 

\subsection{Goals}

A goal for this area is to develop ways of evaluating AI-based simulation methods.  To some extent this will build on existing methods of evaluating any simulation or modeling system.   However, as more AI systems become black boxes in terms of the algorithms and training data, it becomes more essential to develop methods of evaluating their sensitivity and robustness.   In particular, as AI systems can be increasingly used as a part of new modeling techniques (e.g., \cite{jackson2026modeling}) it becomes increasingly important to understand how inputs and prompts to AI affect its output. 
Another important goal relates to the fact that AI-aided research is increasing the speed and quantity of scientific production.  This results in overload in systems built for a different scientific era.  Sorting through AI-generated science will undoubtedly require AI assistance, and producing those systems is beginning, but naturally lags behind the emerging content.  A goal of ensuring that higher quantity science does not become lower quality will require an understanding of AI behavior and systems that can evaluate output in new ways (again, leaning on theme 1).

\section{Theme 3: the Science of Human-AI Interaction}
\label{sec:interaction}

Our third area of discussion concerns the fact that the extensive deployment of AI means that most social and economic systems are no longer simply human systems within the confines of some organization or institutions, but are now systems of interacting humans and AI agents.
Behavioral sciences have long studied human interactions, including how they are shaped by the various institutions and organizations humans have developed. We are now faced with a world in which humans are not just interacting with each other and through institutions, but are interacting with various forms and layers of AI. AI guides the information that people get and with whom they interact.  It can influence how people vote, how they do their jobs, how they educate themselves, and what they do in their spare time. 
AI Behavioral Science includes the study of complex and multi-layered interactions between humans and AI in a variety of contexts.  This involves modeling and understanding such interactions, as well as 
taking a normative approach of envisioning the future of human-AI collaboration to maximize benefits for society \citep{vaccaro2024combinations}.

\begin{figure}[h!]
    \centering
    \includegraphics[width=\linewidth]{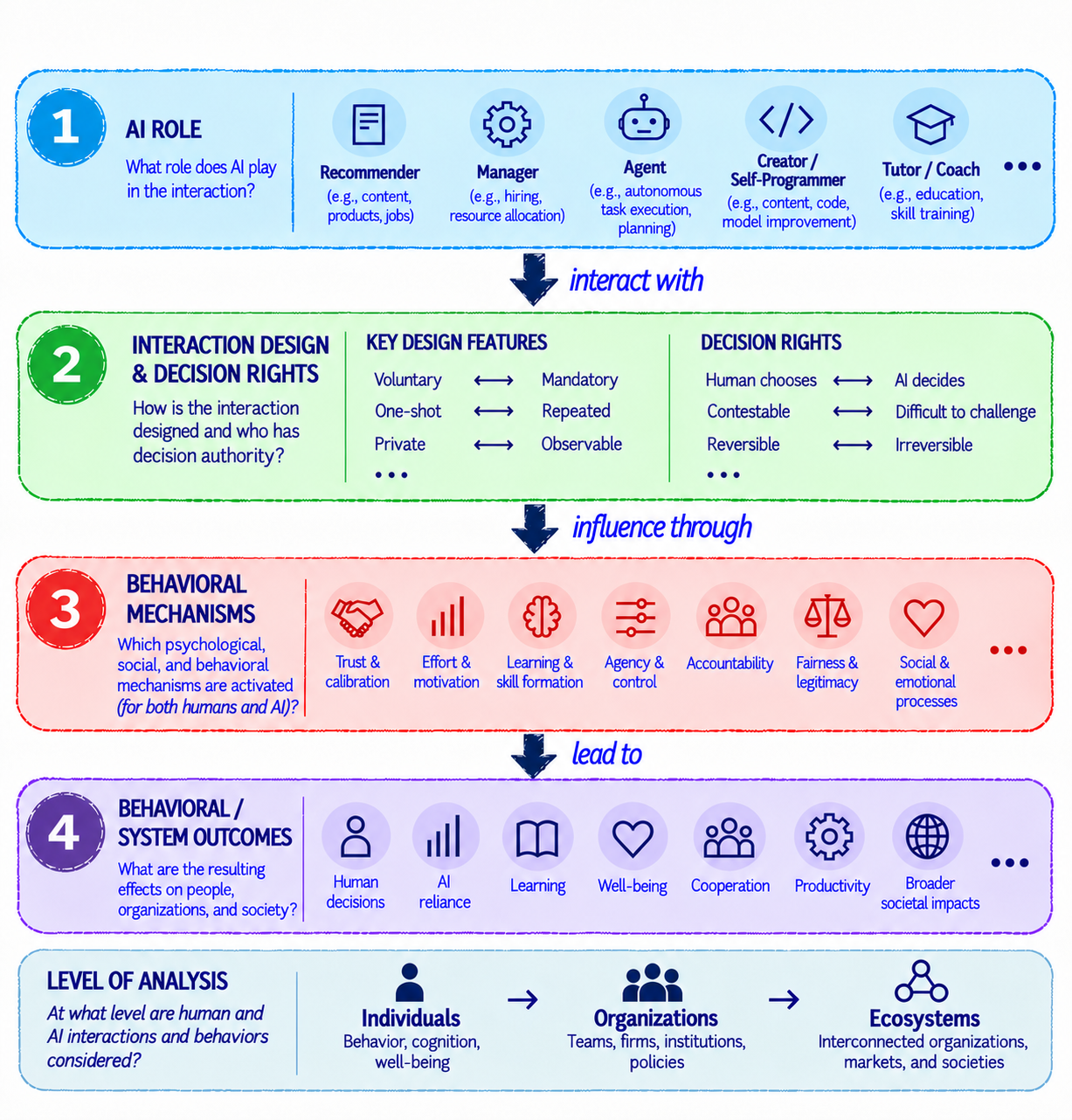}
    \caption{Human–AI interaction framework. AI roles interact with the design of human–AI interactions and the allocation of decision rights, influencing behavioral mechanisms and resulting outcomes across individual, organizational, and ecosystem levels. }
    \label{fig:human-ai-framework}
\end{figure}

Human–AI interactions can be organized along several dimensions that provide a common framework to study their behavioral consequences (Figure \ref{fig:human-ai-framework}). AI can occupy roles ranging from recommender and collaborator to tutor, manager, or companion. The effects of these roles depend on how the interaction is designed and how decision rights are allocated between humans and AI. These features can activate behavioral mechanisms, including trust and calibration, effort and motivation, learning and skill formation, and agency and control, which, in turn, shape outcomes ranging from individual decisions and well-being to organizational productivity and broader societal impacts. Importantly, these effects operate at multiple, interacting levels: from individuals and organizations to entire ecosystems that encompass economies and societies.

We use this framework to organize the examples and issues that follow, while recognizing that particular applications may involve multiple roles, mechanisms, outcomes, and levels simultaneously. The framework applies across three interconnected levels of analysis: individual human–AI interactions; organizational interactions within teams, firms, and institutions; and ecosystem-level interactions across interconnected organizations, markets, economies, and societies. 

\subsection{Three Levels of AI-Human Interactions}

In Section \ref{sec:assess}, our discussion was AI-centric.  Once AI is embedded in some social or economic context and interacting with humans and other AI agents, these interactions can change both human and AI behavior. Understanding these effects requires moving beyond the analysis of individual agents in isolation to examine how interactions and their consequences unfold across multiple levels. 

\subsubsection{Micro-Level Human-AI Interactions}

We begin with a discussion of how AI and humans each behave when interacting, essentially as a pair.  At this level, the framework highlights how the role assigned to AI and the design of the interaction shape behavior through mechanisms such as trust, learning, motivation, and agency. In a sense, this expands game theory and behavioral economics to include agents whose payoffs and objective functions are quite different from the motivations generally presumed in human-to-human interactions.   

One form of interaction of this type is a rapidly emerging and ubiquitous role of AI as an advisor, information provider, and aid.  That is, AI is being used in helping humans in various ways: by providing information, helping them complete tasks, and helping them ``improve'' behaviors.

Starting from the last, the AI-enabled nudges and personalized interventions discussed in Section \ref{sec:protocols} offer opportunities to promote positive behavioral change\citep{gabriel2024misinfoeval}. 
Advanced data analytics and multimodal behavioral insights can enable more accurate bias detection than traditional methods. For example, AI-based tools have shown promise in detecting cognitive biases in text, with preliminary evaluations reporting performance above that of the average human evaluator \citep{atreides2024cognitive}.  
Meanwhile, AI-powered decision systems can holistically address multiple cognitive biases simultaneously and optimize decisions across domains \citep{celiktutan2024people,costello2024durably} - a fundamental shift from traditional decision-by-decision nudging. Crucially, AI enables dynamic personalization—moving beyond one-size-fits-all nudges to tailor interventions based on individual heterogeneity, fostering deeper collaboration between behavioral science and digital innovation for scalable impact \citep{collins2024building,burton2024large}.

AI also shows promise in providing emotional support, especially in areas like mental health and education \citep{yin2024ai, opel2026transforming}. An important feature of these interactions is how users perceive AI relative to human counterparts. In contexts where people fear stigma or judgment, some users may perceive AI as less judgmental and therefore be more willing to disclose sensitive information or seek support. AI systems can also provide low-cost and readily accessible support at times when human assistance may not be available. These potential benefits should not be confused with AI being inherently unbiased or empathetic: its responses can reflect biases in training data and system design, and perceived empathy need not correspond to human empathy. Challenges lie in preventing users from becoming overly emotionally dependent, as well as ensuring AI responses promote healthier human behaviors rather than reinforcing stigmas and stereotypes \citep{moore2025expressing}. 
Understanding the dynamics of these interactions requires new research on how humans form beliefs and mental models about the beliefs and mental models AIs have of them \citep{madarasz2023projective}. Along with the potential benefits, there are also concerns that AI could lead to loss of skills, reduce learning, or lead to emotional dependence or addiction \citep{naseer2025psychological,acemoglu2026ai}.

AI has also been reshaping how people learn social and other skills by enabling interactive, low-risk environments for practicing complex skills. Especially with social simulation, AI can be customized to imitate a variety of specific individuals and idiosyncratic behaviors. This opens up possibilities for people to learn in low-stakes artificial contexts where they interact with AI rather than other humans. The AI Partner and AI Mentor framework \citep{yang2024social} proposed using AI for social skill training by using simulated practice for experiential learning and by using simulated AI mentors to offer personalized feedback based on domain expertise and factual knowledge.  Here, AI-based simulation allows for repeated, adaptive, and individualized experiences, which are especially valuable in domains that involve communication and interpersonal dynamics.

In education, social learning, and professional training, AI systems are increasingly used to simulate learners or clients, provide real-time guidance, and create low-risk environments for practice \citep{markel2023gpteach, wang2024tutor, shaikh2024rehearsal, argyle2023leveraging, louie2024roleplay, hsu2025helping}. These applications illustrate a shift from AI as an information provider toward a tutor and collaborator, with repeated and adaptive interactions that affect learning, effort, skill formation, and agency.

All these cases reveal that AI can change how people learn, which raises profound research questions that may reshape educational theory and practice for decades to come \citep{zhang2021ai,McGee2025}. Key questions include when AI assistance improves rather than impairs cognitive development \citep{lehmann2024ai,bastani2025}, whether existing learning theories extend to AI-mediated instruction \citep{Li2025Is}, and how human pedagogical expertise and AI capabilities can best be combined \citep{garg2024impact}.

Here, AI increasingly shifts from information provider to tutor and collaborator, with repeated and adaptive interaction affecting learning, effort, skill formation, and human agency.

As the transformative potential of generative AI reshapes the breadth and depth of human-AI interaction, understanding the psychological and social consequences of these interactions on human participants has become critically important. Beyond the extensive efforts that highlight productivity and creativity gains from AI assistance \citep{chen2024large,brynjolfsson2025generative}, recent findings highlight concerning patterns on human well-being, e.g. high-intensity users demonstrate increased emotional dependence and reduced socialization \citep{phang2025investigating}. In addition to usage volume, interaction modes can significantly impact human perceptions of themselves: when LLMs initiate human-AI co-writing processes, users report diminished perceived autonomy and self-efficacy compared to human-first approaches where humans create initial drafts before requesting LLM revisions \citep{Ren2025Order}. These early signals suggest research could expand beyond performance metrics to develop nuanced frameworks for healthy and sustainable human-AI relationships that preserve, if not enhance, psychological well-being, social connection, and human autonomy. We anticipate increased investigation into how interaction design, usage patterns, and contextual factors might foster beneficial rather than detrimental psychological outcomes as these systems become increasingly embedded in daily life.

It is already clear that humans act differently when interacting with AI than with other humans \citep{shechtman2003media,mou2017media,gnewuch2024more,goergen2025ai} though humans may tend to evaluate AIs using the same frameworks at times\citep{dreyfussraux}. On the one hand, AI may be thought of as being less judgmental than other humans. On the other hand, it might also be less capable of being empathetic in particular situations where other humans may have ample experience. There are also questions of confidentiality and security, as we already know things that are shared on social media can become widely public quickly. Understanding the evolving interactions and how they change how humans behave\citep{almog2026human,almog2026when}, and when they can lead to overall improvements in behavior, is an important frontier \citep{caplin2025abcs,chan2006,khatua2026cooperbench}. There are many different arenas in which the role of AI will be different, so understanding how the context in question impacts the emerging behavior is vital. These are areas in which the modeling of games and other complex systems in the behavioral sciences can be leveraged\citep{dreyfusshoong}.  

Despite the enthusiasm about human-AI complementarity, recent literature also highlights several potential concerns. For example, worker well-being can erode when AI decision assistants become ``digital overseers,'' raising stress and diminishing perceived autonomy \citep{jarrahi2023algorithmic,cram2022examining}. As many AI models are trained with the objective of defeating human players or achieving superhuman performance \citep{silver2017mastering}, their behavior in real-world scenarios may set humans up for failure. In scenarios like autonomous driving, the AI often turns the human into a last-resort fallback who must intervene only during rare, dangerous edge cases \citep{hancock2019future,weaver2022systematic}. In fact, even perfect hand-offs cannot solve fundamental communication and persuasion limits, and the inefficiency of processing such information may lead to human operators routinely discounting or over-trusting the algorithms \citep{green2019principles,carton2020feature}. Recent empirical results and developments in impossibility theorems have highlighted these “no-free-lunch” limits of AI interpretability or human-AI complementarity \citep{doshi2017towards,bansal2021does,donahue2022human,peng2025no,acemoglu2026ai}.

Evidence from hiring reveals two enduring lessons about human-AI collaboration.  First, well-designed algorithms can raise efficiency and equity: recommender systems shorten time-to-fill vacancies \citep{horton2017effects}, algorithmic interview filters increase acceptance rates and subsequent productivity \citep{cowgill2020bias}, and AI-generated candidate scores help recruiters advance women in STEM \citep{avery2024does}.   AI's function can help in recruitment (candidate sourcing) and selection (candidate evaluation)  \citep{kochling2020discriminated}, and Will et al. \cite{will2023people} find that AI generally equals or exceeds human recruiters in terms of efficiency and effectiveness.  Second, human behaviors impact whether those gains materialize: managers who discount machine advice make worse hires \citep{hoffman2018discretion}, while AI drafting tools induce more, but lower-quality, job posts without better matches \citep{wiles2024more}.  
In a natural field experiment, \citep{Jabarian2025Voice} provide evidence on how real-time voice AI agents reshape incentives, bias, and complementarities in labor markets. 

These examples illustrate how the same broad role for AI, as recommender or manager, can produce different outcomes depending on the allocation of decision rights and how humans calibrate their reliance on algorithmic advice.

A growing body of work demonstrates how AI's capabilities can be applied across much wider domains
of human-AI interactions. 
The potential of human-AI interaction is emerging in the context of education, science, interviews, {groups and experts}, managing risk, law enforcement, medicine, and creative arts \citep{tabrizi2025magic,goldfarb2025patterns,tabrizi2025behavioral}. It has been shown that LLM dialogues can reduce conspiracy theorizing \citep{costello2024durably}, and LLMs can facilitate compromise in highly polarized groups \citep{tessler2024ai}.

Together, these examples underscore the need to assess AI not only in isolation, but within the interactions it creates—where roles, decision rights, and behavioral responses jointly determine outcomes.

\subsubsection{Organizations and AI-Human Ecosystems}

 The framework also applies at the organizational level, where multiple humans and AI agents interact within teams, firms, and institutions. At this level, the roles AI plays, the design of interactions, and the allocation of decision rights can shape coordination, accountability, and collective outcomes. Understanding these interactions requires examining feedback effects, externalities, and synergies that may not be apparent when individual human–AI interactions are studied in isolation.

For example, people may use AI to gather information but then interact on an AI-guided platform, which shapes team production. Analyzing each part in isolation can miss important interactive features, externalities, inefficiencies, and weaknesses due to systemic properties. The study of such interactions may involve the development of new game theory and other behavioral science tools.  It is also proving to be useful to use combinations of AI together, much as humans might use teams whose members have different expertise to solve a problem (e.g., \cite{park2025maporl}).  Multi-member teams of AI can police each other, query each other, refine each other's perspectives, and do other things that help provide improved collective behavior.  

In particular, the basic idea of ``assessing AI'' as a single entity can overlook interactions among agents that matter for organizational performance and welfare. In organizational settings, different AI agents with different designs and capabilities can take different roles, and the alignment between these AI agents, as well as with humans with whom they interact, is essential. This raises questions about the design of such systems and the allocation of responsibilities and decision rights between humans and AI.   

Biases in behaviors in any one component may be amplified when combined with other components.  There may also be unanticipated interactions and feedback between AI agents performing different functions.  We already see this in scientific research, where AI is used to help conduct research, aid in writing research, and evaluate and referee research.  Knowing that research may be evaluated by AI can feed back to change the way the research is conducted and written up, and may even lead to changes in which projects researchers choose to pursue and how much effort they devote to any given project. This illustrates how AI adoption in one organizational function can influence behavior and outcomes in others. 

These organizational interactions also extend beyond the boundaries of individual teams, firms, and institutions, giving rise to broader human-AI ecosystems, which we discuss next. 

\subsubsection{Ecosystem-Level Human–AI Interactions}

At the ecosystem level, human-AI interactions extend across organizations, markets, economies, and societies. The framework's dimensions remain relevant, but their application becomes more complex: AI agents may occupy different roles across interconnected systems, decision rights may be distributed among firms, users, and regulators, and feedback effects can propagate across organizational and market boundaries. Understanding these interactions requires examining how individual and organizational behaviors aggregate into broader economic, political, and societal outcomes.

For example, this is already evident in rapidly changing search and question-answering systems, central to our lives, where AI agents may be involved in ranking, query formation, and document generation. The alignment of these agents is crucial for overall system performance and social welfare \citep{nachimovsky2025multi}. More generally, AI ecosystems may consist of producers, consumers, and market makers, each involving different forms of AI or human agents. Aligning these components and allocating responsibilities and decision rights can be crucial to ecosystem behavior. 

At this level, the acceleration in AI’s emergence gives rise to a need for (new) insights on how AI will transform whole economies and societies. Behavioral macroeconomics has matured in recent decades to incorporate human motivations that explain departures of the real economy from the competitive general equilibrium model \citep{akerlof2002}. Now we need insights from AI Behavioral Science about the behavior of AI and how humans use AI to inform new macro-level models. 

As mentioned above, AI’s capabilities lead many to predict that they will surpass humans at any purely cognitive task in the near future, with the remainder of tasks eventually to follow. We need to use the methods outlined in Section~\ref{sec:assess} along with new ones to assess whether these predictions are borne out. Forms of AI have already displaced labor in a variety of arenas \citep{brynjolfsson2025canaries}, from accountants to travel agents. It has dramatically altered the retail shopping industry and is poised to do the same in the driving/transportation industry.  It shows promise in new areas, for instance, dramatically outperforming analysts in picking stock portfolios \citep{dehaan2025shadow}. This does not necessarily mean that AI will replace humans in all economic activities, especially given psychological and sociological barriers to adoption\citep{Almog2025}, but it does suggest the potential for large declines in the relative importance of human labor in the production of goods and information and across much wider parts of the economy than previous technological transformations \citep{idetalamas2025,acemoglu2026ai}. Some of the psychological and sociological barriers have been explored earlier in this section and more research is needed to enumerate these barriers and understand how they are changing over time. 

At least for the medium term, the economic concept of {\sl comparative advantage} provides many insights.  Even if one mode of production (non-human) has an absolute advantage on all or most tasks, as long as it is not free, prices and wages adjust so that the less-productive mode (human labor) is still used where it has the best relative advantage. Tasks that are least amenable to the list of AI capabilities above, for instance, are those that are constantly changing and presenting new challenges, require dexterity, etc. These face the least pressure for displacement. 
Historical technological transitions illustrate how productivity gains can coincide with substantial labor reallocation and uneven distributional effects \cite{manyika2017jobs,acemoglu2022tasks,brynjolfsson2022turing,eloundou2024gpts,aghion2025different}. Mechanization dramatically reduced agriculture’s share of employment over the twentieth century, while manufacturing automation since the 1980s similarly displaced some workers even as it increased productivity and created new tasks and occupations 
\citep{bls}.
These experiences do not necessarily predict the effects of AI, but they underscore the importance of studying both aggregate gains and their distribution across workers, firms, and communities \citep{eloundou2024gpts, weintraub2025generative}.

There are many views about the macroeconomic impact of the newest wave of technologies including AI and various forms of robotics and automation \citep{acemoglu2018artificial,jackson2026economic,pissarides2025labour}. A model focused solely on the automation potential of current AI finds it will contribute less than one percentage point to economic growth over the next decade \citep{acemoglu2025simple}, while models emphasizing positive feedback between AI investments and the ease of future scientific innovations anticipate a large transformation of the economy, with much larger global growth rates \citep{aghion2017artificial,abis2024changing,erdil2025gate}.

While accurately predicting technological development and deployment over decades may be impossible, models for scenario planning can help guide policy as uncertainty is resolved \citep{bommasani2025advancing}.  For example, ongoing research builds on agent-based computable general equilibrium overlapping generations models to understand the interaction of government tax, trade, and regulatory policy with technological innovations in creating macroeconomic and distributional outcomes \citep{benzell2021simulating,benzell2024simulating}.  AI tools will also be useful in improving and accelerating the development of these models: neoclassical agents may be replaced with AI ones which have more realistic behaviors and can handle strategic situations; surrogate AI models of the simulation will speed model development and testing; and AI-based prediction tools can be married to structural projections to retain the best features of both.  

Existing economic models, including general-equilibrium and agent-based approaches, can incorporate externalities, heterogeneous agents, and strategic interactions. However, modeling human-AI ecosystems presents a distinct challenge: we have limited understanding of how human decisions change when interacting with AI, whether and how AI agents adapt their behavior to humans and other AI agents, and what collective dynamics emerge from interactions among multiple human and AI agents. Developing and validating models that capture these behavioral responses and their system-level consequences remains an important research challenge. 

Anticipating disruptions and identifying occupations and regions that will be most severely impacted \citep{andreadis2025local} can help steer government fiscal, regulatory, and education policy and minimize the damage of such `creative destruction' and {guide the development and deployment of AI to the most beneficial paths}. For example, policymakers must avoid the ``Turing Trap'': the economic pitfall that arises when innovators focus too heavily on building human-like AIs that substitute for workers rather than augmenting them. Investment in AI designed entirely to replace humans can widen wage inequality and leave vast reserves of untapped productivity on the table. Redirecting incentives toward complementary, human-enhancing AI can help society escape the trap and ensure that the gains from AI are widely shared \citep{brynjolfsson2022turing}.  There is evidence that new AI technologies are doing both (e.g., \cite{zhou2020effect,hoffmann2024generative,choi2025human}).
This also presents political challenges, as competitive economies lack incentives to introduce complementary rather than substituting technologies.

Equally important are the political implications of AI and automation. Acemoglu et al. model the trade-off between the redistribution, regulation of automation, and repression, finding a complementarity between automation and repression \citep{acemoglu2026automation}. Insights into AI behavior and how humans are using AI in the political domain continue to inform debates about institutional regulation of AI, the influence of AI on the social media and electoral landscape, the impact of AI on statecraft, and many other urgent political economy issues \citep{agrawal2026political}. 

An important challenge is to predict which parts of the economy and political life, if any, might be left relatively untouched in the AI transformation. As with previous technological shocks, there is a danger of over-extrapolating effects from some sectors to others or from some institutions to others.  AI will also result in further technological advances that both complement and compete with humans, and can alter communication and political interactions. This is why we need principled models of the mechanisms through which AI likely will or will not alter the tradeoffs inherent in many economic and political decisions and to be clear-eyed about bottlenecks. 

Finally, we briefly mention that it is important to understand the actual production of AI itself \citep{frank2019toward,jackson2026economic}.  The brevity of this discussion is not because of a lack of importance, but because of the rapid emergence of this as a topic.  Traditional industrial organization studies \citep{tirole1988theory,varian2018artificial} offer substantial insight, but the details of this new industry are complex and unique. The potential profits from various applications are enormous, which is attracting massive amounts of investment.  Much of the development is by private parties who wish to keep the design details private. Given that the designers’ objectives and those of society may not align (e.g., designing systems that maximize human engagement and emotional response rather than learning and welfare), it is essential to understand the economics, politics, profitability, incentives, legality, and ethics of AI production \citep{kearns2019ethical,stahl2024ethics}.  The economies of scope and scale, as well as the network externalities involved with AI, can lead to inefficient market-based outcomes \citep{acemoglu2021harms}. Technology is emerging at a much faster rate than we can comprehend, and across various governmental jurisdictions, making this another area in urgent need of study, and may require governmental intervention and politics on an international scale. 

\subsection{Goals}

At the individual level, a central goal is to understand how AI roles, interaction design, and decision rights shape human behavior through mechanisms such as trust, learning, motivation, and agency, and to identify when human–AI interactions improve rather than diminish human outcomes and well-being. 

At the organizational level, the goal expands to understanding how multiple human and AI agents interact within teams, firms, and institutions, including coordination, feedback effects, and the allocation of decision rights. Developing behavioral models and empirical methods capable of measuring these interactions and their organizational consequences will be essential. A model of human and AI agents with varying competence and coordination abilities \citep{imas2026organizing}, generates different predictions about the ecosystem depending on the time horizon. In the short to medium run, increases in AI competence change the ratio of humans to AI agents within firms and reduce firm size. In the long run, if AI coordination improves enough, organizations may transition to larger AI-dominated firms. These predictions can be tested against observed changes in firm size, workforce composition, and AI adoption over time.  Other models of such dynamics are also needed so that we better understand the relationship between assumed context and dynamic predictions. Empirically, where platform-wide AI integration occurs, one can study whether AI adoption leads to better performance (e.g., monthly sales) or simply changes in costs, and the mechanisms through which  changes happen \citep{barwick2026ai}. It is feasible to document which organizations gain and whether AI expands a market rather than merely redistributing among firms.

At the ecosystem level, the goal extends to understanding how human-AI and AI-AI interactions propagate across interconnected organizations, markets, economies, and societies, including externalities, alignment problems, and emergent collective dynamics. The overarching motivation is to understand how these interactions affect social welfare in a world that increasingly intertwines human and AI behavior.  The challenge is that increasing overall productivity is not synonymous with increasing welfare.  This potential divergence has always been an issue, as any new technology has the potential to create winners and losers and can increase inequality, but the potential for increased inequality with AI deployment is larger than ever.  Assessing welfare therefore requires examining not only aggregate productivity but also employment, wages, inequality, and human well-being, identifying which workers, firms, and communities gain or lose, and distinguishing short-term disruptions from longer-term outcomes \citep{jackson2026economic}.  

As \cite{brynjolfsson2022turing} and others have made clear, a goal is to have AI complement human abilities as much as possible, rather than only substituting for human production.
That can help alleviate the dangers of increased inequality.  This loosely stated overarching goal then requires more specific research objectives across the three levels: understanding human–AI team production and how it depends on the setting, understanding the incentives of industries to produce and adopt AI, and understanding how humans naturally tend to use AI.

\section{Closing Remarks}

Science is not always efficient. 
It involves synergies and complementarities between different researchers, teams, and the public and private sectors. It also provides public goods, as the generated knowledge can be shared broadly. Given these features, neither markets nor a set of disparate private and public institutions, such as universities, naturally lead to efficient levels of scientific discovery, nor to efficient interaction across different areas of study.   
As ``AI Behavioral Science'' is emerging rapidly and organically, the purpose of this article has been to
synthesize an area of study and to provide agendas for study that reinforce and support each other. 
The three themes identify key areas of focus, which new explorations are needed in each and how they fit together, as well as associated key goals.

A synthesis of those key goals is as follows. For AI assessment and guidance, it is crucial to develop quantifiable methods of evaluating AI's capabilities and potential behaviors, especially given the difficulties in anticipating all the ways in which AI agents might behave in different domains, and especially when they interact with humans and have the potential to dramatically change economic and political organizations and behaviors (Themes 1 and 3). A related priority is to develop principles and guidelines for AI design and deployment with better understanding of AI's behavior (Themes 1 and 2). To use AI effectively as a scientific tool, new ways of evaluating AI-based simulation methods for sensitivity and robustness must be developed (Theme 2).  Similarly, AI behavior must be better understood to design AI-aided systems that can better evaluate the increasing volume of scientific and other output (Theme 2).  AI also has the potential to open new techniques for modeling and understanding human behavior and human-AI interactions and can thus be used to understand the impact of AI on society (Themes 2 and 3).

\newpage
\bibliographystyle{plainnat}
\bibliography{references}  

\newpage
\appendix

\end{document}